\documentclass[prb,twocolumn,floatfix,showpacs,superscriptaddress]{revtex4-1}
\usepackage{graphicx,epsfig,color}
\usepackage[%
    colorlinks=true,
    pdfborder={0 0 0},
    allcolors=blue,
    breaklinks=true
]{hyperref}

\usepackage{threeparttable}
\usepackage{breakcites}
\usepackage[utf8]{inputenc}
\begin{document}

\title{Theoretical investigation of the role of the organic cation in methylammonium lead iodide perovskite using a different method}

\author{Veysel \c{C}elik}\email{vcelik@siirt.edu.tr}
\affiliation{Department of Mathematics and Science Education, Siirt University, Siirt 56100, Turkey}

\begin{abstract}
The hybrid halide perovskite CH$_3$NH$_3$PbI$_3$ is easy to manufacture and inexpensive. Despite these, its efficiency as a solar cell is comparable to today's efficient solar cells. For these reasons, it is attracting a lot of attention today. However, the effects of the CH$_3$NH$_3^+$ (MA) molecule in the perovskite structure on the electronic and structural properties are still a matter of debate. Previous studies have generally focused on the rotation of the MA molecule. In this study,  from a different perspective, the effects of the movement of the MA molecule along the C-N axis are investigated. With this method, the effects of the MA molecule were examined in a more controlled way. In this study, density functional theory (DFT) that accounts for van der Waals (vdW) interactions was used in the calculations for the cases.  According to the data obtained, H-I ionic bonds are formed between the MA molecule and the inorganic framework. Within the structure, the H-I bond length tends to be preserved, although the position of the MA changes. In this mechanism, the I ion plays an important role by moving away from its place in the Pb-I-Pb alignment. The position of the I ion determines the nature of the band gap transition. Another effect is on the value of the band gap. Depending on the position of the I ion, the band gap may narrow by about 0.26 eV. The separation of the I ion from the Pb-I-Pb alignment by the effect of the MA molecule breaks the inverse symmetry. According to the data obtained from this study, this mechanism in the band gap is due to the breaking of the inverse symmetry in the crystal structure.
\end{abstract}

\pacs{68.43.Bc, 68.43.Fg}

\maketitle
                                    
\section{Introduction}
Today, the need for electrical energy is increasing day by day. Solar cells are one of the important technologies for generating electricity without polluting the world. However, cost, production difficulty, and efficiency are all important factors to be considered. In this sense, organic-inorganic perovskites attract a lot of attention today. This is because they are inexpensive to produce \citep{Pablo2014,AntoniettaLoi2013} and their efficiency approaches that of high-efficiency solar cells such as GaAs and CdTe.\citep{ZHANG202018}   In 2009, \citet{Kojima2009} tried to use the lead-based organic-inorganic hybrid perovskite, such as methylammonium lead iodide (CH$_3$NH$_3$PbI$_3$), to make solar cells.  As the data obtained from that study was promising, CH$_3$NH$_3$PbI$_3$ attracted a lot of attention from researchers in the following years. This interest is due to the low defect density\citep{Dong2015} with a long charge carrier lifetime\citep{Bi2016} and diffusion length,\citep{Qingfeng2015} although the material is solution-processable.The power conversion efficiency achieved in 2020 is about 26\%.\citep{Min2021} The CH$_3$NH$_3$PbI$_3$ has shown the fastest improvement in solar cell efficiency of all known materials. The CH$_3$NH$_3$PbI$_3$ has an important place among hybrid organic perovskites. 

Although it is produced by the solution-processable method, its comparable properties to photovoltaic materials produced under special conditions can be partially explained by the behavior of native defects. According to the theoretical work of \citet{Yin2014}, point defects with low formation energy in CH$_3$NH$_3$PbI$_3$ produce shallow defect levels. On the other hand, defects that cause deep energy levels, which act as trap levels, have higher formation energies. However, these data are not sufficient to explain the high efficiency of the material. In addition, the effects of the CH$_3$NH$_3^+$ (MA) molecule should be examined. The effect of the MA molecule on the electronic and structural properties of the material is still a matter of debate. Most theoretical studies report that the MA molecule has no significant effect on the electronic structure of the valence band maximum (VBM) and conduction band minimum (CBM).\cite{Brivio2014,Umari2014} The general view is that the role of the molecule is to stabilize the perovskite structure electrostatically. However, there have been studies showing that the effect of the MA molecule is not limited to this. In their theoretical study, \citet{Motta2015} investigated the effects of the rotation of the MA molecule in the center of the perovskite structure and reported that if it is oriented along a direction like [011], the PbI$_6$ octahedral lattice will be distorted and the band gap will become indirect.  In their theoretical study, \citet{Ong2019} reported that the value and nature of the band gap change depending on the orientation of the MA molecule. 

The temperature has a significant effect on the physical properties of the CH$_3$NH$_3$PbI$_3$. \citet{Quarti2016} reported that optical properties of the CH$_3$NH$_3$PbI$_3$ change gradually when the temperature is increased from 270 K to 400 K. This situation can be associated with the transition from the tetragonal phase to the cubic phase, but the transition to the cubic phase occurs at 327.4 K.\cite{Poglitsch1987} It may be insufficient to explain this situation with a phase transition. Experimental studies show that the motion of the MA molecule increases with the increase in temperature. Experimental work by \citet{Wasylishen1985} revealed that the cation exhibits complete disorientation in the cubic phase at high temperatures but freezes at low temperatures.  The effect of the MA molecule, whose motion increases as the temperature increases, should be investigated further.

The aim of this study is to reveal the role of the MA molecule in the MAPbI$_3$ structure with a more controlled method. In the DFT code I used in this study, the tests I made in the calculations that included the van der Waals distribution correction showed that the initial position of the MA molecule in the structure did not change much during the structural optimization process, and more I ions were positioned relative to the MA molecule. This finding also shows that the initial position of the MA molecule, which has weak interaction with the inorganic frame, is also important for DFT calculations in which van der Waals distribution corrections are included. The idea that this may help explain the role of the MA molecule in a more controlled manner forms the basis of this study. In previous studies on the role of the MA molecule, the structural changes caused by the rotation of the MA molecule which is in center of the PbI inorganic framework and the electronic properties caused by these changes were investigated.\cite{Motta2015,Klinkla2018} In this study,  using a different method, the effects of the movement of the MA molecule along the C-N axis are investigated. With this method, the effects of the MA molecule were examined in a more controlled manner. This method also sheds light on the effect of rotation of the molecule on electronic structure and structural properties.

\section{Computational Method}
The density functional theory (DFT) calculations have been performed based on the projector-augmented wave (PAW)\cite{Blochl,Kresse2} method as implemented in the Vienna ab-initio simulation package (VASP).\cite{Kresse1,Kresse3}  For comparison purposes in the calculations, nine functionals were used for the exchange-correlation energy(E$_{xc}$). Two of these are the Perdew–Burke–Ernzerhof (PBE)\cite{PBE1996} and PBE revised version for solid (PBEsol)\cite{psol} functionals. The others are exchange-correlation functionals within the generalized gradient approximation (GGA)\cite{GGA} and plus van der Waals (vdW) dispersion correction.\cite{Dion2004,Klime2009,Klime2011,Hamada2014,Peng2016} In these functionals, the formula for the exchange-correlation energy is of the form 
\begin{equation}
E_{\rm xc}=E_{\rm x}^{\rm GGA}+E_{\rm c}^{\rm LDA}+E_{\rm c}^{\rm nl}\,,
\end{equation}
where the exchange energy $E_{\rm x}^{\rm GGA}$ uses the revised PBE functional (revPBE),  $E_{\rm c}^{\rm LDA}$ is the local density approximation (LDA) to the correlation energy. 
The $E_{\rm c}^{\rm nl}$ is the non-local energy term that approximates the consequences of non-local electron correlation. Due to the effort to obtain experimental data, different versions of these functionals have been developed. In this study, to determine and compare the characteristic properties of the most thoroughly studied CH$_3$NH$_3$PbI$_3$ perovskite structure, vdW-DF\cite{Dion2004}, vdW-DF2, rev-vdW-DF2\cite{Hamada2014}, SCAN+rVV10\cite{Peng2016} and the functional groups developed by Klime\ifmmode \check{s}\else \v{s}\fi{} \textit{et al}.\cite{Klime2009,Klime2011} (optPBE-vdW, optB88-vdW, and optB86b-vdW) were used. 

In this work, 12-ions cubic unit cells are used for cases.  Temperature can change the phase structures of CH$_3$NH$_3$PbI$_3$.\cite{Poglitsch1987,Baikie2013,Lopez2020} Perovskite CH$_3$NH$_3$PbI$_3$ is in an orthorhombic (Pnma) phase below approximately 161 K, while at temperatures above approximately 327 K, it transforms from a tetragonal (I4/mcm) phase to a cubic phase (Pm$\bar{3}$m). The unitcell parameters are related to the cubic phase parameter as $\sqrt{2}a_c\times\sqrt{2}a_c\times2a_c$ and $\sqrt{2}a_c\times2a_c\times\sqrt{2}a_c$ for I4/mcm and Pnma, respectively. 

For geometry optimizations and density of states (DOS), the Brillouin zones were sampled with 6$\times$6$\times$6 and 12$\times$12$\times$12 Monkhorst-Pack\cite{mp} $k$-point grids, respectively. Plane wave basis set was used to expand the wavefunctions up to a kinetic energy cutoff value of 520 eV. The fine FFT grids with high precision settings were used throughout the calculations. Atomic positions and cell parameters were optimized until residual forces were below 0.01 eV/\AA. The structures in this study were fully optimized.

Bader analysis based on the atom-in-molecule (AIM) theory was utilized for qualitatively describing interatomic charge distributions. By integrating Bader volumes around atomic sites, local charge depletion/accumulation may be estimated. These volumes are sections of the real space cell bounded by zero-flux surfaces of the gradient vector field of charge density. In this work, using a grid-based decomposition technique created by Henkelman's team,\cite{Henkelman} the charge states of atomic species were determined (Table~\ref{table3}).

\section{Results \& Discussion}
There are weak bonds between the organic cation and the inorganic framework in hybrid organic-inorganic perovskites. Considering this situation, functionals using vdW dispersion correction were used in this study. A comparison was made to find the most accurate result among these functionals. The comparison of the results obtained is listed in Table~\ref{table1}. In the structure used for comparison, the direction of the C-N bond of the organic molecule is [100], and the cation molecule was placed in the center of the inorganic lattice. In all calculations, the structure was fully optimized without restrictions. 

\begin{table}[h]

\begin{threeparttable}
\caption{The comparison of computational and experimental data for CH$_3$NH$_3$PbI$_3$.
Latice parameters and band gaps are in angstroms and eV, respectively.}
\label{table1}
\begin{ruledtabular}
\begin{tabular}{lccccc}
{Functional} &  a   &   b  &  c   &{Band gap} & Ref. \\[1mm] \hline
LDA   &  - & - & - & 1.46 & \citenum{Brivio2013} \\
PBE   & 6.48 & 6.46 & 6.47 & 1.59 & This work\\
PBE & 6.46 & - & - & 1.65 & \citenum{Yu2016}\\
PBEsol   & 6.30 & 6.26 & 6.28 & 1.36 & This work\\
PBEsol  & 6.29 & - & - & 1.35 & \citenum{Yu2016}\\
HSE   & 6.40 & 6.38 & 6.39 & 2.15 & This work \\
HSE+SOC   & - & - & - & 1.14 & \citenum{yin2015}\\
DFT-1/2& 6.17 & 6.15 & 6.22 & 2.96 & \citenum{Tao2017}\\
DFT-1/2+SOC& 6.17 & 6.15 & 6.22 & 1.81 & \citenum{Tao2017}\\
vdW-DF & 6.61 & 6.48 & 6.51 & 2.03& This work\\
vdW-DF2  & 6.62 & 6.48 & 6.52 & 1.96 & This work\\
optPBE-vdW  & 6.48 & 6.38 & 6.40 & 1.85 & This work\\
optB88-vdW  & 6.40 & 6.31 & 6.33 & 1.76 & This work\\
optB86b-vdW & 6.36 & 6.30 & 6.31 & 1.66 & This work\\
rev-vdW-DF2   & 6.38 & 6.29 & 6.31 & 1.72 & This work\\
Experimental  & 6.329\tnote{a} & 6.329\tnote{a} & 6.329\tnote{a}& 1.60\tnote{b}  \\
\end{tabular}
\end{ruledtabular}
\begin{tablenotes}       
        \item[a] Reference\citenum{Poglitsch1987}.
        \item[b] Reference\citenum{Lopez2020}.
    \end{tablenotes}
\end{threeparttable}
\end{table}

One of the important data that should be obtained correctly in the calculation of the electronic structure of semiconductors is the value of the band gap. As can be seen in Table \ref{table1}, the value of the band gap can be calculated very close to the experimental data with calculations using PBE. However, PBE overestimates the value of lattice parameters. On the other hand, the PBEsol functional underestimates the band gap and the HSE06 functional\cite{HSE}, which is a method for correcting the exchange energy\cite{CELIK2021}, overestimates the band gap. It is a well-known error that standard DFT underestimates the band gap of semiconductors. However, as can be seen in Table~\ref{table1}, especially the PBE estimates, the band gap is quite compatible with the experimental data. The reason for this is the relativistic effect of heavy Pb and I ions in the structure. One of the results supporting this is that the calculated band gap decreases by about 1 eV when the effect of spin-orbit coupling (SOC) is included in the calculations.\cite{Wang2020,Tao2017} It can be used to correct the error of functionals that overestimate the band gap. The obtained band gap value is 2.15 eV and 2.96 eV\cite{Tao2017} for HSE06 and DFT-1/2, respectively. When SOC is included, it becomes 1.14 eV\cite{yin2015} and 1.81 eV\cite{Tao2017} for HSE06 and DFT-1/2, respectively. However, the experimental band gap value is about 1.60 eV.\cite{Lopez2020} Since the defect levels in the electronic structure were not investigated in this study, SOC was not included in the calculations.  On the other hand, it was reported in the previous study that SOC did not significantly affect the ground state properties of this material.\citep{Umari2014} According to the results obtained in this study, the optB88, optB86b and rev-vdW-DF2 functionals produce good results for CH$_3$NH$_3$PbI$_3$ when compared to other functionals in Table~\ref{table1}. When these three functionals are compared, the optB86b functional gives closer results to the experimental data. Therefore, the calculations were made using the optB86b functional.

\begin{figure}[h]
    \centering
    \includegraphics[width=0.5\textwidth]{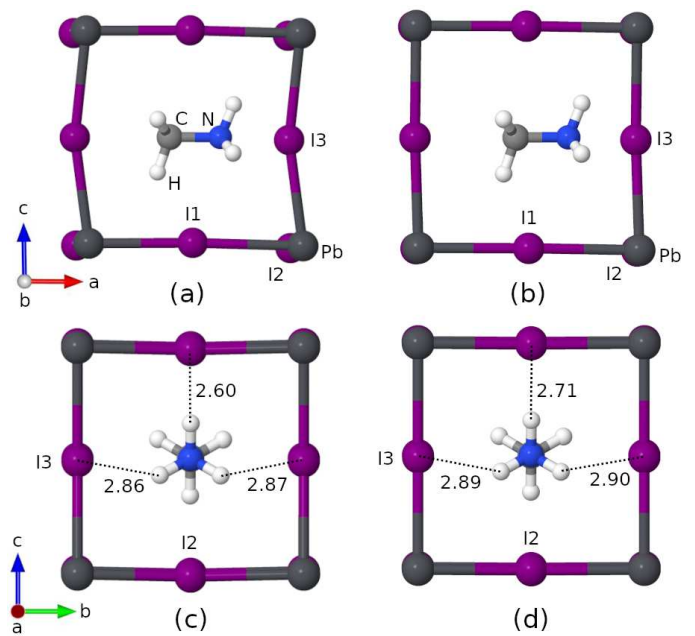}
    \caption{Two different directions of fully optimized structures in the S0 and S5 cases. Since cases S0 and S5 are critical according to Fig.~\ref{fig2}, two structures are shown in the figure. The H-I bond length for other cases is shown in Table~\ref{table2}. (a) and (c) show two different directions for case S0. (b) and (d) show two different directions for case S5.}
    \label{fig1}
\end{figure}

In order to determine the stable orientation of the molecule, calculations were made in cases where the C-N bond of the MA molecule in the center was oriented [100], [110] and [111]. The total energies of the cases where the orientation of the molecule is [100] and [110] are very close to each other. The total energies of these two cases are approximately 0.033 eV lower than the total energy of the [111] case. The absence of a significant energy difference between the orientations indicates that the orientation may be temperature sensitive. This is in agreement with the experimental data.\cite{Wasylishen1985}
 
\begin{table*}[t]
\begin{ruledtabular}
\caption{For the cases studied, the lattice parameters, H-I bond lengths, band gap (Eg), and the nature of the transition in the band gap are listed.The MA position indicates where the MA molecule is placed in the initial structure and its unit is the angstrom (\r{A}). A value of 0.000 indicates that the MA molecule is placed in the center of the inorganic lattice. Other values show how much the MA molecule is shifted in the [100] direction relative to the 0.000 position in the initial structure.}
\label{table2}
\begin{tabular}{lcccccccc}
Case & MA Position & a & b & c & H-I2 & H-I3 & E$_g$ (eV) &  Transition \\ [1mm] \hline
S0 & 0.000 & 6.35 & 6.29 & 6.30	& 2.60 & 2.87 & 1.66  & I\\
S1 & 0.063 & 6.35 & 6.29 & 6.31 & 2.65 & 2.87&	1.59 & I \\
S2 & 0.126 & 6.33 & 6.31 & 6.32 & 2.65 & 2.89 & 1.51  & I \\
S3 & 0.189 & 6.33 & 6.32 & 6.33 & 2.67 & 2.90  & 1.47 & D\\
S4 & 0.252 & 6.32 & 6.32 & 6.32 & 2.70 & 2.90  &	1.42 & D\\
S5 & 0.315 & 6.33& 6.33 & 6.33  & 2.71 & 2.90  & 1.41  & D \\
S6 & 0.378 & 6.32 & 6.33 & 6.33  & 2.67 & 2.87 & 1.40  & D \\
S7 & 0.441 & 6.24 & 6.36 & 6.36	 & 2.62 & 2.84& 1.42  & D  \\
\end{tabular}
\end{ruledtabular}
\end{table*}

\begin{table*}[t]
\begin{ruledtabular}
\caption{Bader analysis of the studied cases. H1, H2 and H3 tags represent hydrogen ions bonded with C ion, while H4, H5 and H6 tags represent hydrogen ions bonded with N ion. The I ions represented by the I1, I2 and I3 tags are shown in Fig.~\ref{fig1}.}
\label{table3}
\begin{tabular}{lcccccccccccc}
Case & C & N & H1  & H2 & H3 & H4 & H5 & H6 & Pb & I1 & I2 & I3 \\ [1mm] \hline
S0 & +0.197 & -1.097 & +0.055 & +0.105 & +0.109 & +0.465 & +0.461 & +0.463 & +0.895 & -0.549 & -0.550 & -0.553 \\
S1 & +0.177 & -1.140 & +0.099 & +0.109 & +0.107 & +0.471 & +0.468 & +0.466 & +0.890 &-0.545 &-0.550 & -0.553\\
S2 & +0.168 & -1.170 & +0.097 & +0.110 & +0.108 & +0.487 & +0.489 & +0.472 & +0.884 &-0.546 &-0.551 & -0.547\\
S3 & +0.151 & -1.145 & +0.100 & +0.110 & +0.114 & +0.473 & +0.492 & +0.468 & +0.885 &-0.549 &-0.555 & -0.543\\
S4 & +0.204 & -1.141 & +0.102 & +0.077 & +0.114 & +0.472 & +0.466 & +0.471 & +0.893 &-0.547 &-0.564 & -0.548\\
S5 & +0.166 & -1.151 & +0.101 & +0.117 & +0.115 & +0.474 & +0.464 & +0.480 & +0.895 &-0.548 &-0.562 & -0.549\\
S6 & +0.165 & -1.154 & +0.100 & +0.117 & +0.116 & +0.472 & +0.469 & +0.476 & +0.890 &-0.544 &-0.560 & -0.546\\
S7 & +0.162 & -1.160 & +0.099 & +0.118 & +0.116 & +0.477 & +0.472 & +0.471 & +0.885 &-0.541 &-0.560 & -0.542\\
\end{tabular}
\end{ruledtabular}
\end{table*}

The value of the lattice parameters of the initial structures used for the calculations is 6.30 \r{A}. This value corresponds to an average value when the literature is reviewed. As I mentioned before, the total energies of the [100] and [110] cases are very close to each other. The orientation of the C-N bond is [100] in all of the initial structures used in the calculations. In line with this choice, \citet{Quarti2016} found that the [001] direction is the most stable direction for MAPbI$_3$, which has a symmetrical structure. 

For comparison purposes, eight initial structures were used in the calculations. In the first of these structures, the MA molecule was placed in the center of the inorganic lattice. In the other structures, the MA molecule was shifted from the center in the [100] direction with equal intervals. For eight structures, the distances of MA molecule from the center are 0.000 \r{A}, 0.063 \r{A}, 0.126 \r{A}, 0.189 \r{A}, 0.252 \r{A}, 0.315 \r{A}, 0.378 \r{A} and 0.441 \r{A}. These initial structures were labeled as S0, S1, S2, S3, S4, S5, S6 and S7, respectively (Table~\ref{table2}). The structures were fully optimized without any restrictions.

\begin{figure}[h]
    \centering
    \includegraphics[width=0.45\textwidth,trim={0 0 0 0},clip]{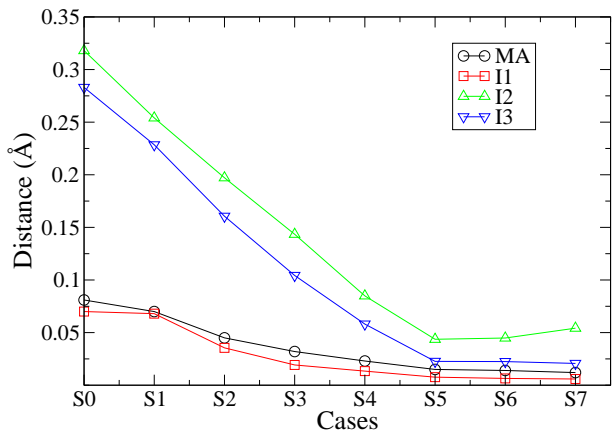}
    \caption{The distances of MA molecule and I ions from their initial coordinates in the [100] direction when the structures are optimized. The decrease in the values in the graph shows the movement in the [$\overline{1}$00] direction, while the increases indicate the movement in the [100] direction. I1, I2 and I3 ions are shown in Fig.~\ref{fig1}.}
    \label{fig2}
\end{figure}

As can be seen in Table ~\ref{table2}, the values of the lattice parameters a, b and c vary according to the position of the MA molecule. The value of the lattice parameters of the cases S3, S4, S5 and S6 is quite close to the experimental value of 6.329 \r{A}(ref.\citenum{Poglitsch1987}). At the same time, distortion in the cubic structure decreases as the MA molecule approaches the inorganic framework up to a certain position.  However, as with the S7 case, distortion begins when the MA molecule approaches the inorganic lattice more than a certain distance. In the S7 case, the I2 ion is pushed outward. In all cases, there was no remarkable change in the coordinates of the Pb ion during the optimization processes.  The dominant movement of the ions is in the [100] direction. The H-I bonds formed in S0 and S5 cases are shown in Fig.~\ref{fig1}. The length of the bonds between the I and H ions in all the optimized structures is very close to each other (Table ~\ref{table2}). The results show that the structure tries to keep the H-I bonds at a certain value, and as can be seen in Fig.~\ref{fig2}, I ions play the biggest role here. However, the MA molecule does not displace as much as the I ions during the optimization process. Here, it is the MA molecule that determines the location of the I ions. These findings also clearly show how the rotation of the MA molecule affects the structure.

Table~\ref{table3} shows the results of the Bader analysis of ions for the cases examined. In the cases studied, the charges of the MA molecule and the PbI inorganic frame are approximately +0.76\textit{e} and -0.76\textit{e}, respectively. As can be seen in Table~\ref{table3}, in the S0 case, the charge of each of the H ions bonding with the N ion in the MA molecule decreased by about 0.46e and became positively charged. In contrast, the charges of each of the I ions in the PbI frame increased by about 0.55e and became negatively charged. The Pb ion is positively charged. This shows that H-I ionic bonds are formed between the inorganic frame and the MA molecule. When the DOS graph and Bader analysis results for the inorganic lattice are examined, it can be concluded that the bond between Pb and I ions is both ionic and covalent. There is no significant difference between the data of the S0 case and the other cases in Table~\ref{table3}. The charge distributions are similar.

\begin{figure*}[]
    \centering
    \includegraphics[angle=0, width=0.85\textwidth]{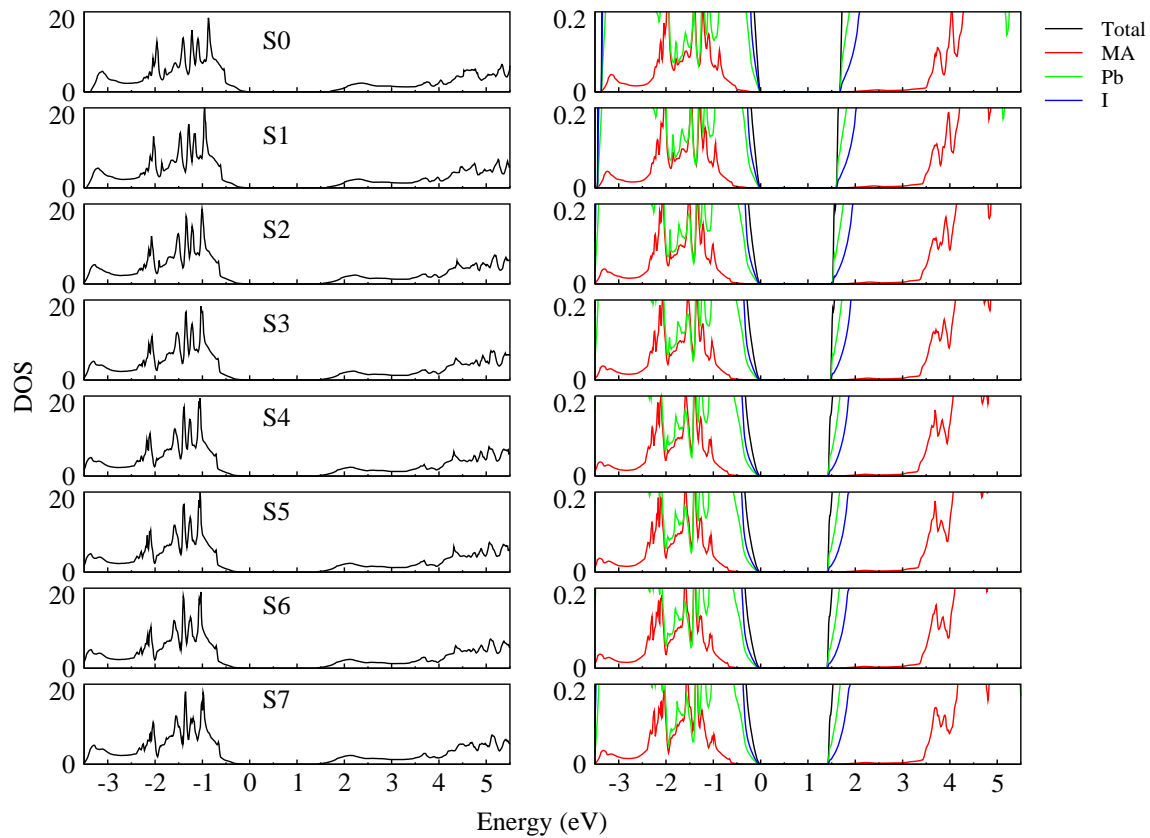}
    \caption{Calculated DOS plots for the cases studied. The graphs on the left show the total DOS of the cases. The graphs on the right show the partial DOS of the ions and the molecule. The band edges are detailed in the graph on the right. }
    \label{fig3}
\end{figure*}

The density of states (DOS) pattern obtained for the eight cases of the MAPbI$_3$ is shown Fig.~\ref{fig3}. The total DOS patterns of all structures are similar except for the band gap. The I p energy levels dominate in the valence band in all cases of the MAPbI$_3$ structure, while Pb p energy levels dominate in the conduction band. In the band edges, the energy states of the MA molecule are very low compared to Pb and I ions. Therefore, in order to show the contribution of the MA molecule to the electronic structure, the band edges are detailed in Fig.~\ref{fig3}(b). As can be seen in the DOS pattern shown in Fig.~\ref{fig3}(b), the MA energy states in all structures are below the valence band maximum (VBM) and above the conduction band minimum (CBM). In all structures, the location of the MA states relative to the CBM and VBM states is approximately the same. In the all cases, the MA energy states are approximately 0.2 eV below the VBM and 0.7 eV above the CBM. On the other hand, both VBM and CBM edges are composed of Pb-I hybrid energy states. These data indicate that the effect of the MA molecule on the band edges is indirect. 

\begin{figure}[]
    \centering
    \includegraphics[width=0.45\textwidth,trim={0 0 0 0},clip]{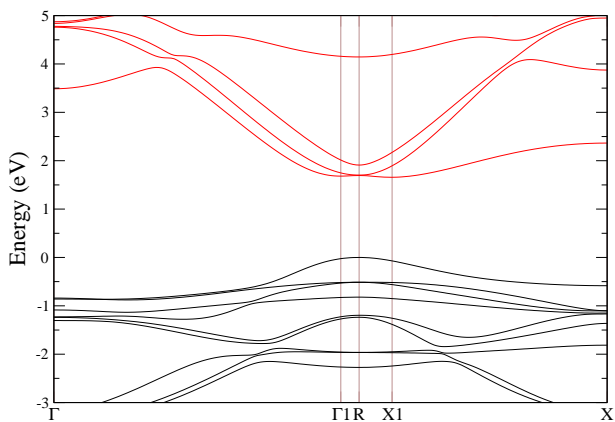}
    \caption{Detailed band structure for case S0. }
    \label{fig4}
\end{figure}

\begin{figure*}[]
    \centering
    \includegraphics[width=0.85\textwidth]{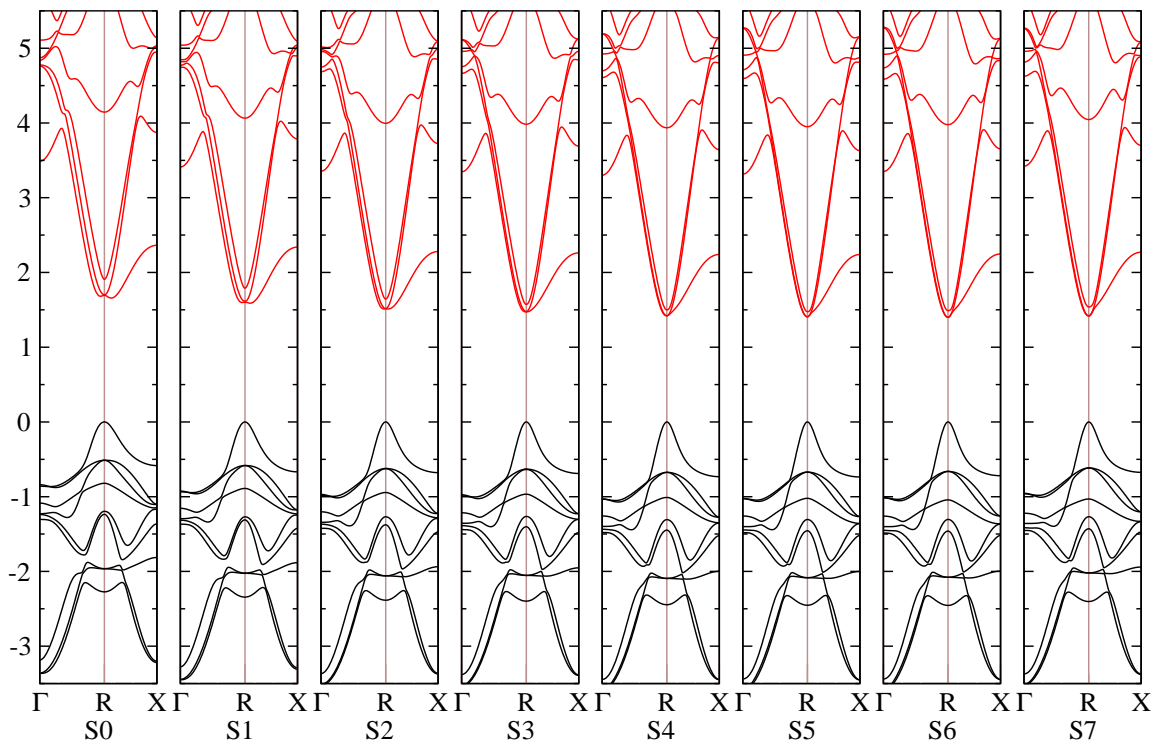}
    \caption{Band structures calculated for the cases studied. }
    \label{fig5}
\end{figure*}

The plot obtained to investigate the effects of MA molecule on the electronic band structure of S0 case is shown in Fig.~\ref{fig4}. The plots show the regions where the band gap is the lowest. In the case of S0, the nature of the band gap is indirect and its value is 1.66 eV. The minimum values of the CBM are around the R symmetry point, and the lowest level is in the R$\rightarrow$X1 direction. Compared to the CBM at R point, the difference between them is about 0.035 eV. This value decreases towards case S7. On the other hand, a lower energy level occurs in the R$\rightarrow\Gamma$1 direction than the R point. However, the level here is not as low energy as in the R$\rightarrow$X1 direction. As can be seen in Fig.~\ref{fig5}, these two low CB levels around R converge at point R as MA approaches the inorganic frame. When the cases are compared, from S0 to S7, there is tapering towards the band gap in the CBM at the R symmetry point.  At the same time, tapering towards the band gap occurs at the R point in VBM, but this is not as pronounced as in CBM. As a result of this process, in the case of S3, the transition nature of the band gap changes and becomes direct. The transitional nature of the band gap is also direct in structures after the S3 case (Table ~\ref{table2}). At the same time, the band gap narrows from S0 to S6. The difference between the band gaps of the S0 and S6 structures is about 0.26 eV. This decrease in band gap is consistent with the work of \citet{Ong2019} where they showed that the rotation of the MA molecule decreases the band gap by about 0.2 eV in the [111] orientation relative to the [001] orientation. Another result of their work is that the rotation of the molecule defines the direct and indirect nature of the energy band gap. The compatibility of the data obtained from this study with the data obtained from the study conducted for the effect of rotation of the MA molecule shows that the method in this study can also explain the effects of the rotation of the MA molecule. When the data in Fig.~\ref{fig2} and Fig.~\ref{fig5} are combined, it becomes clear that the position of the I ion plays an important role in these changes in CBM. In the S3 case, the I2 and I3 ions shown in Fig.~\ref{fig2} have moved away from their initial positions by 0.145 Å and 0.105 Å in the [$\overline{1}$00] direction, respectively. These values may be critical values for the position of the I ion. As can be seen in Fig.~\ref{fig2}, these values decrease after the S3 case. As the MA molecule continues to be brought closer to the organic framework, the I2 ion starts to move slightly in the [100] direction after the S5 case. However, in the case of S7, the nature of the band gap is direct, since the position of the I2 ion is below the critical level.

The situations where the nature of the transition in the band gap is indirect show characteristics similar to the Rashba-type effect.\citep{Rashba1960} Rashba-type effect describes the splitting of spin angular momentum in electronic systems in the presence of SOC and the absence of inverse symmetry. With the SOC effect, spin degeneration in the k-space is lifted and the VBM and/or CBM shift away from the symmetry points in the Brillouin region. However, as I mentioned before, SOC was not included in the calculations in this study. But it has been reported that this effect continues when SOC is included in the previous study.\cite{Motta2015} On the other hand, the inclusion of SOC in the calculations makes this effect more pronounced. The possible reason for this is the presence of heavy atoms such as Pb and I in the structure. However, the changes in the nature of the band gap shown in this study are due to the structural changes caused by the MA molecule by attracting the I ions. In the S0 case, the displacement of the I ions from the Pb-I-Pb alignment to the MA molecule breaks the inverse symmetry. As the I ions return to the Pb-I-Pb alignment after the S3 case, the inverse symmetry in the structure begins to occur again and the nature of the band gap becomes direct from the S3 case. Accordingly, the findings from this study show that the main source of this mechanism in the band structure is the break in inverse symmetry. As can be seen in Fig.~\ref{fig5}, this mechanism is more effective in CBM than in VBM. This mechanism in the band structure can affect the charge carrier dynamics. Light absorption can occur via direct transitions, but recombination of cooled carriers may be hindered by band shift in k-space, resulting in a forbidden indirect transition.\citep{Zheng2015}

\section{CONCLUSION}

According to the results obtained from this study, the location of the MA molecule in the initial structure is important in the calculations using the optB86b-vdW functional. During the optimization process, the position of the MA molecule does not change much, and the changes are mostly in the PbI framework. The MA molecule is bonded to the PbI framework by H-I ionic bonds. In the structure, the H ions attached to the N ion of the molecule are more positively charged than the other ions, and the I ions are more negatively charged. The H-I bond length in all cases is very close to each other. This shows that within the structure, the H-I bond length tends to be conserved with respect to the position of the MA molecule. The I ions play the biggest role in this mechanism and moves away from its initial position within certain limits. This movement of the I ion directly affects the electronic structure. The effect of the MA molecule on the electronic structure is indirect, which is also supported by the DOS pattern. 

The position of the I ion directly affects the electronic band structure. When the I ion moves away from its place in the octahedral structure at certain distances under the influence of the MA molecule, the nature of the transition in the band gap changes from direct to indirect. The main reason for this mechanism in the band gap is the breaking of the inverse symmetry in the crystal structure. Although this mechanism is similar to the Rashba-type effect, SOC was not included in the calculations in this study.  However, as I mentioned before, when SOC is included in the calculations, the effect of this mechanism occurring in the band gap becomes more apparent.\cite{Motta2015} The main reason for this is the effect of Pb and I heavy ions in the crystal structure.

Another finding is that the location of the I ion in the crystal affects the band gap. From the S0 case to the S6 case, the band gap narrows by about 0.26 eV. This narrowing is proportional to the increase in the symmetry of the Pb-centered octahedral structure in the crystal. The results obtained from this study also shed light on the effects of the rotation of the MA molecule. The results obtained from this study agree with the studies conducted to investigate the effects of the rotation of the MA molecule.

According to the data obtained from this study, the electronic and structural properties of MAPbI$_3$ can be controlled by moving the MA molecule closer to and away from the inorganic framework. This controllable mechanism could be important for the technological applications of MAPbI$_3$.


\bibliography{references}

\end{document}